\documentclass[sigconf]{acmart}

\AtBeginDocument{%
  }

\usepackage{multirow}
\usepackage{array}
\usepackage{makecell}
\usepackage{subfigure}

\newcommand{\loss}{\mathcal{L}}

\copyrightyear{2023}
\acmYear{2023}
\setcopyright{rightsretained}
\acmConference[SIGIR '23]{Proceedings of the 46th International ACM SIGIR
Conference on Research and Development in Information Retrieval}{July 23--27,
2023}{Taipei, Taiwan}
\acmBooktitle{Proceedings of the 46th International ACM SIGIR Conference on
Research and Development in Information Retrieval (SIGIR '23), July 23--27,
2023, Taipei, Taiwan}
\acmDOI{XXXXXXX.XXXXXXX}
\makeatletter
\gdef\@copyrightpermission{
  \begin{minipage}{0.3\columnwidth}
   \href{https://creativecommons.org/licenses/by/4.0/}{\includegraphics[width=0.90\textwidth]{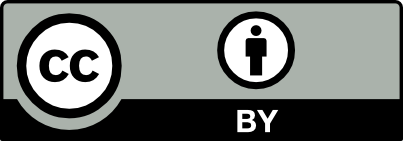}}
  \end{minipage}\hfill
  \begin{minipage}{0.7\columnwidth}
   \href{https://creativecommons.org/licenses/by/4.0/}{This work is licensed under a Creative Commons Attribution International 4.0 License.}
  \end{minipage}
  \vspace{5pt}
}
\makeatother

\begin{document}

%%
%% The "title" command has an optional parameter,
%% allowing the author to define a "short title" to be used in page headers.
\title{PiTL: Cross-modal Retrieval with Weakly-supervised Vision-language Pre-training via Prompting}

%%
%% The "author" command and its associated commands are used to define
%% the authors and their affiliations.
%% Of note is the shared affiliation of the first two authors, and the
%% "authornote" and "authornotemark" commands
%% used to denote shared contribution to the research.

\author{Zixin Guo}
\affiliation{%
  \institution{Aalto University}
  \city{Espoo}
  \country{Finland}}
\email{zixin.guo@aalto.fi}

\author{Tzu-Jui Julius Wang}
\affiliation{%
  \institution{Aalto University}
  \city{Espoo}
  \country{Finland}}
\email{tzu-jui.wang@aalto.fi}

\author{Selen Pehlivan}
\affiliation{%
  \institution{Aalto University}
  \city{Espoo}
  \country{Finland}}
\email{selen.pehlivantort@aalto.fi}

\author{Abduljalil Radman}
\affiliation{%
  \institution{Aalto University}
  \city{Espoo}
  \country{Finland}}
\email{abduljalil.saif@aalto.fi}

\author{Jorma Laaksonen}
\affiliation{%
  \institution{Aalto University}
  \city{Espoo}
  \country{Finland}}
\email{jorma.laaksonen@aalto.fi}

%%
%% By default, the full list of authors will be used in the page
%% headers. Often, this list is too long, and will overlap
%% other information printed in the page headers. This command allows
%% the author to define a more concise list
%% of authors' names for this purpose.

\renewcommand{\shortauthors}{Zixin Guo, Tzu-Jui Julius Wang, Selen Pehlivan, Abduljalil Radman, \& Jorma Laaksonen}
%% No italics

%%
%% The abstract is a short summary of the work to be presented in the
%% article.
\begin{abstract}
Vision-language (VL) Pre-training (VLP) has shown to well generalize VL models over a wide range of VL downstream tasks, especially for cross-modal retrieval. However, it hinges on a huge amount of image-text pairs, which requires tedious and costly curation. On the contrary, \textit{weakly-supervised} VLP (W-VLP) \cite{wang2023learning} explores means with object tags generated by a pre-trained object detector (OD) from images. Yet, they still require paired information, i.e. images and object-level annotations, as supervision to train an OD.

To further reduce the amount of supervision, we propose Prompts-in-The-Loop (PiTL) that prompts knowledge from large language models (LLMs) to describe images. Concretely, given a category label of an image, e.g. \textit{refinery}, the knowledge, e.g. \textit{a refinery could be seen with large storage tanks, pipework, and ...}, extracted by LLMs is used as the language counterpart. The knowledge supplements, e.g. the common relations among entities most likely appearing in a scene.  
We create IN14K, a new VL dataset of 9M images and 1M descriptions of 14K categories from ImageNet21K \cite{deng2009imagenet} with PiTL.
Empirically, the VL models pre-trained with PiTL-generated pairs are strongly favored over other W-VLP works on image-to-text (I2T) and text-to-image (T2I) retrieval tasks, with less supervision. 
The results reveal the effectiveness of PiTL-generated pairs for VLP. 
\end{abstract}

%%
%% The code below is generated by the tool at http://dl.acm.org/ccs.cfm.
%% Please copy and paste the code instead of the example below.
%%


%%
%% Keywords. The author(s) should pick words that accurately describe
%% the work being presented. Separate the keywords with commas.
%% A "teaser" image appears between the author and affiliation
%% information and the body of the document, and typically spans the
%% page.
% \begin{teaserfigure}
%   \includegraphics[width=\textwidth]{sampleteaser}
%   \caption{Seattle Mariners at Spring Training, 2010.}
%   \Description{Enjoying the baseball game from the third-base
%   seats. Ichiro Suzuki preparing to bat.}
%   \label{fig:teaser}
% \end{teaserfigure}

% \received{20 February 2007}
% \received[revised]{12 March 2009}
% \received[accepted]{5 June 2009}

%%
%% This command processes the author and affiliation and title
%% information and builds the first part of the formatted document.
\maketitle

\begin{figure}[t!]
    \centering
    \subfigure[VLP: Vision-language Pre-training. W-VLP: Weakly-supervised VLP.]{
    \includegraphics[width=0.4\textwidth]{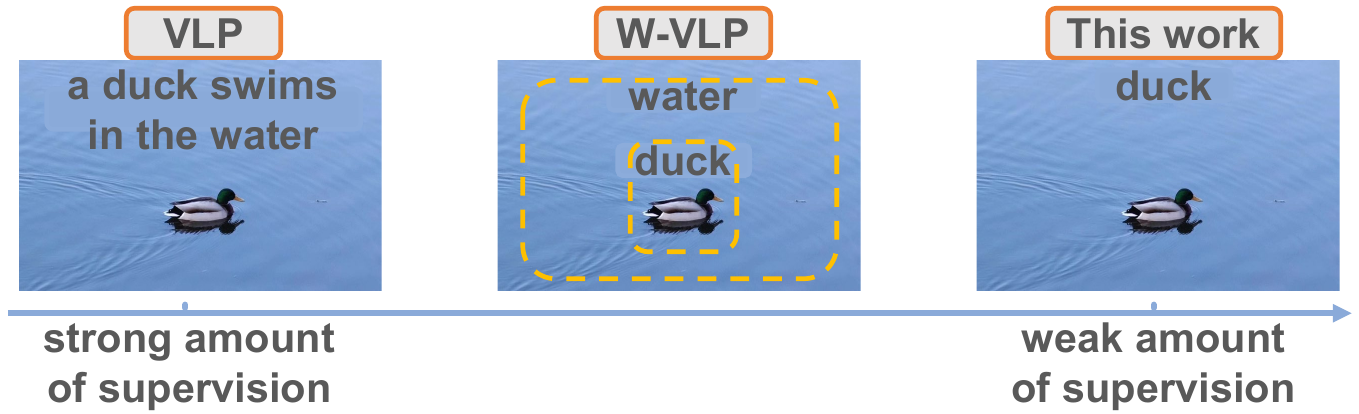}}
      
    \subfigure[Our proposed Prompts-in-The-Loop (PiTL).]{
		\includegraphics[width=0.4\textwidth]{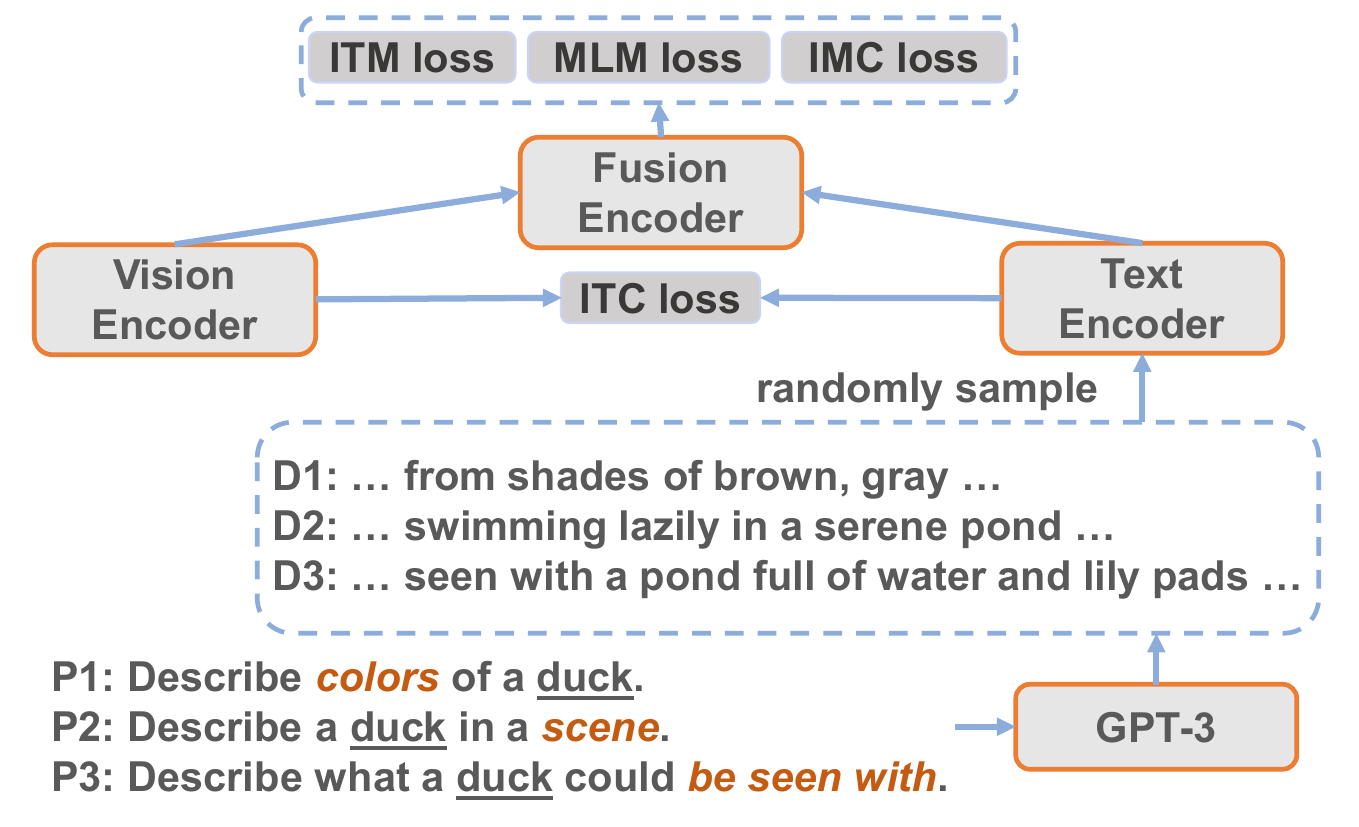}}
    
    \vspace{-0.3cm}
    \caption{(a) Different VLP problem settings. The \textit{de facto} VLP requires the aligned images and the corresponding descriptions. W-VLP learns on multiple object-level annotations, such as the object bounding boxes and the labels. Our proposed W-VLP, Prompts-in-The-Loop (PiTL), comes with the least supervision from the image category labels. (b) PiTL leverages the shared prompts (P1-3) on the category label and the responses (D1-D3) from a large language model to generate image-description pairs. Please refer to Table~\ref{table:prompt_example} for the nine prompts used in this work.}
    
    \label{fig:teaser}
\end{figure}

\section{Introduction}
Vision-language (VL) models have been advancing rapidly with the introduction of various Vision-language Pre-training (VLP) methods. The models for VLP can adapt to various downstream tasks involving VL modalities, such as cross-modal retrieval~\cite{radford2021learning,luo2022clip4clip,chen2020uniter,kim2021vilt,jia2021scaling}, visual question answering~\cite{zhang2021vinvl,byun2022grit,li2020oscar,kim2021vilt,gan2020large}, visual captioning~\cite{li2021unimo,li2022unimo,li2022blip,li2023blip,guo2022clip4idc}, etc. 
The secret recipe of a VLP model comes with (1) a decent amount of \textit{webly-supervised}\footnote{Some of the gathered image-text pairs may not be highly relevant as they are not validated by a human.} image-text pairs, (2) effective pre-training objectives which learn cross-modal interaction, and (3) sufficient resources, e.g. GPUs/TPUs, which enable large-scale training. Those models can also scale well with more image-text data, i.e. the models excel better in downstream tasks with more pre-training data \cite{radford2021learning,jia2021scaling,li2021align,singh2022flava,zellers2021merlot,alayrac2022flamingo,li2022blip,li2023blip,li2022mplug,wang2022image,li2022unimo,yang2022vision}.

\subsection{Weakly-supervised VLP}
While the success of VLP methods relying on huge amounts of image-text annotations has been proven, 
a less visited research path is emerging to pursue more data-efficient VLP. The data-efficiency of a VL model can be viewed from the amount of supervision. That is, would a VL model pre-trained with less image-text data remain as performant in downstream tasks? This question leads to works~\cite{li2020unsupervised,wang2022vlmixer,zhou2022unsupervised,chen2022end,wang2023learning} in \textit{weakly-supervised} VLP (W-VLP)
%, such as U-VisualBERT \cite{li2020unsupervised}, VLMixer \cite{wang2022vlmixer}, $\mu$-VLA \cite{zhou2022unsupervised}, E2E-UVLP \cite{chen2022end}, and WFH \cite{wang2023learning}. 
that aims at \textbf{not} relying on image-text pairs from, e.g. SBU Captions \cite{ordonez2011im2text} and Conceptual Captions (CC) \cite{sharma2018conceptual,changpinyo2021conceptual}.

Without the aligned images and texts, these works instead resort to a pre-trained object detector (OD) that generates object tags, i.e. the visual entities detected in the given image. The paired images and the object tags offer weaker supervision than those from the image-sentence pairs, but are still effective as the cross-domain bridge. However, training an OD relies on object-level annotations, which is still a form of supervision \cite{wang2023learning}. This seems to deviate these VLP works from the fully unsupervised path, which aims to remove any kind of cross-modal supervision.

\subsection{Towards Unsupervised VLP}
The \textit{unsupervised} VLP (U-VLP), which aims at learning a VL model without any supervision across modalities, remains a daunting challenge. As a step towards U-VLP, we introduce Prompts-in-The-Loop (PiTL) that generates highly effective image-text pairs for W-VLP without an OD. PiTL capitalizes on image-level, i.e. a category label per image, instead of object-level supervision from the object bounding boxes and the corresponding object categories. This leads to a much harder W-VLP setting since much underlying information about an entity could no longer be inferred, such as the common co-occurrence of the visual entities, e.g. a \textit{chair} and a \textit{desk}, in a scene, and the entity relations. e.g. a \textit{person} usually sits on a \textit{bench}.

Specifically, given images with category labels, e.g. a \textit{duck} as shown in Fig.~\ref{fig:teaser}, we prompt large language models (LLMs), e.g. GPT-3 \cite{brown2020language}, to generate descriptions as the external knowledge of the category labels of the images. In fact, different prompts provide different focuses on each target category, e.g. one that emphasizes colors: "\textit{Describe the colors seen from a/an} <\textit{category}>\textit{?}", and another that emphasizes relations with other entities: "\textit{Describe what could a/an} <\textit{category}> \textit{be seen with?}". This encourages a VL model to associate all plausible visual traits, entities, actions, and scenes pertaining to the target categories.

The prompting paradigm is becoming trendy. Recent works \cite{pratt2022does,menon2022visual,yang2022language} reveal that textual prompts generated by LLMs lead to significant improvement in zero-shot image classification with VLP models like CLIP \cite{radford2021learning}. For instance, prompting a VL model with the LLM-generated description, "\textit{Goldfish are small, orange fish with shiny scales.}", is more likely to find matches in the visual domain than the generic "\textit{A photo of a goldfish.}". Likewise, our work explores if the LLM-generated descriptions could be proved useful as well in the W-VLP setting.

Our contributions are summarized as follows. Firstly, we propose PiTL that generates an image-text dataset IN14K containing 9M images with 1M descriptions of 14K categories from the "Winter21" release of ImageNet-21K \cite{deng2009imagenet}. Second, trained with half of the samples in IN14K, our models are shown competitive to the state of the arts in image-to-text (I2T) and text-to-image (T2I) retrieval tasks. With full IN14K, our models significantly outperform them, e.g. on MSCOCO-5K \cite{lin2014microsoft} by 11\% and 10\%, respectively, on I2T and T2I. Moreover, our models are comparable with VLP models trained with 4M aligned image-text pairs from, e.g. CC3M and SBU Captions, etc. Lastly, PiTL does not only come with the least cross-modal supervision among W-VLP works, but also leads to a small gap between the W-VLP and VLP performances.  

\vspace{-6pt}
\section{W-VLP with Image-level Supervision}

VLP aims at learning VL alignments given a large number of image-sentence pairs. The methodology is concluded as (1) learning shared semantics around VL modalities, (2) learning cross-modal context, e.g. masked modeling \cite{devlin2018bert}, and (3) learning to explicitly match images and texts. 
With the same aim, the existing W-VLP methods leverage OD-generated tags to form VL pairs. What is usually neglected is the cost of pre-training such an OD, which usually requires 10+ object-level annotations to be effective. The proposed PiTL aims at relaxing the requirement of having an OD via prompting LLMs~\cite{brown2020language} to generate descriptions for the object categories.

\subsection{Forming Image-text Pairs via Prompting}
PiTL elicits knowledge about an object category from nine prompts of different perspectives with an LLM. Five descriptions are collected for each prompt. Table~\ref{table:prompt_example} summarizes the nine prompts and their focuses. Some of them are more visually-relevant (\textbf{P1-6}), some focus more on knowledge around the target category (\textbf{P7-8}), and some are more open-ended (\textbf{P9}). We study the effectiveness of the descriptions generated by each prompt later in Sec.~\ref{sec:ablations}.

Among PiTL-generated pairs, an image can be paired with different descriptions as long as they are of the same category. In pre-training, the positive pairs for the Image-Text Contrastive and Image-Text Matching losses (introduced later in Sec.~\ref{sec:pt_losses}) are drawn from the images and descriptions of the same categories, and the negative pairs from those of the different categories. In this setup, pre-training with PiTL-generated pairs encourages the VL models to learn cross-modal alignment at the category level, i.e. images of a target category aligned to a group of descriptions, instead of instance level, i.e. an image aligned with a description as in other VLP works. As such, a given image would be associated with the plausible categorical visual traits, entities, actions, and scenes through the VL models.

\begin{table}[t]
\begin{minipage}{\linewidth}
\caption{LLM prompts and focuses for <\textit{category}>, e.g. \textit{duck}. }
\label{table:prompt_example}

\begin{center}

\scalebox{0.7}{\setlength\tabcolsep{6pt}
\begin{tabular}{
p{0.5cm}
p{6cm}<{\raggedright}
p{4cm}<{\raggedright}
}
\toprule
Types &  Prompts  &  Focuses  \\
\midrule
\textbf{P1} & Describe colors of a <\textit{category}> & colors \\ 
\textbf{P2} & Describe shapes of a <\textit{category}> & shapes\\  
\textbf{P3} & Describe textures of a <\textit{category}> & textures\\  
\textbf{P4} & Describe visual appearances of a <\textit{category}> & summarized visual appearances \\  
\textbf{P5} & Describe a <\textit{category}> in a scene & scenes \\  
\textbf{P6} & Describe what a <\textit{category}> could be seen with & relations with other entities\\  
\textbf{P7} & Describe the places a <\textit{category}> has been seen & places\\  
\textbf{P8} & Describe the main activities of a <\textit{category}> & activities \\
\textbf{P9} & Describe what is it like to be a <\textit{category}> & first-person view \\  
\bottomrule
\end{tabular}}
\end{center}

\end{minipage}
\vspace{-0.4cm}
\end{table}

\subsection{VL Model Architecture}

Our model architecture follows a state-of-the-art VL model, ALBEF \cite{li2021align}, which has a multi-modal encoder fusing the representations generated by visual and textual encoders. Indeed, any VL model with image-text inputs could also be used instead, as proposing a new VL architecture is not the focus of this work.

Specifically, given an image $V$ and its paired text description $T$, the vision encoder $g_v(\cdot)$ follows ViT \cite{dosovitskiy2020image} consisting of a 12-layer Transformer that generates the image embedding as $g_v(V)=(x_v^{CLS}, x_v^{1}, \dots, x_v^{n_v})$. The text encoder $g_t(\cdot)$ is a 6-layer Transformer encoder that embeds the input text as $g_t(T)=(x_t^{CLS}, x_t^{1}, \dots, x_t^{n_t})$, where $x^{CLS}_{v}$ and $x^{CLS}_{t}$ are the representations of the $[CLS]$ tokens summarizing the image and the text, respectively. $n_v$ and $n_t$ are the numbers of image patches and textual tokens, respectively. A fusion encoder $g_f(\cdot)$ consisting of a 6-layer Transformer learns the interaction across the VL modalities encoded as $g_v(V)$ and $g_t(T)$, and generates $g_{f}(g_v(V), g_t(T))=(x_f^{CLS}, x_f^{1}, \dots, x_f^{n_t})$.

\subsection{Pre-training Losses}\label{sec:pt_losses}

Our PiTL VL-models are pre-trained with four losses \cite{li2021align,yang2022vision} that all contribute equally to the total loss $\loss$:
%which evenly contribute to the total loss $\loss$:
    \begin{flalign}
        \loss=\loss_{ITC}+\loss_{ITM}+\loss_{MLM}+\loss_{IMC},
    \end{flalign}
where each objective is described as follows.\\
\noindent\textbf{Image-Text Contrastive (ITC)} aims to retain high and low similarities between the positive and negative image-text pairs, respectively. 
To obtain the ITC loss, one first calculates 
    \begin{flalign}
        p_m^{IT}(V)=\frac{{e^{s(V,T_m)/\tau}}}{\sum_{i=1}^M e^{s(V, T_i)/\tau}},\;
        p_m^{TI}(T)=\frac{{e^{s(T,V_m)/\tau}}}{\sum_{i=1}^M e^{s(T, V_i)/\tau}},
    \end{flalign}
where $s(V, T)=(x_v^{CLS})^Tx_t^{CLS}$ measures the dot-product similarity of an image-text pair $(V, T)$. $p_m^{IT}(V)$ and $p_m^{TI}(T)$ are image-to-text and text-to-image similarities, respectively. $\tau$ is a learnable temperature parameter and $M$ is the size of the queues storing the image and textual class embeddings. The ITC loss is then defined as
    \begin{flalign}
        \loss_{ITC}=\frac{1}{2} \mathbb{E}_{V, T \sim D}[H(\textbf{y}^{IT}_V, \textbf{p}^{IT}(V)) + H(\textbf{y}^{TI}_T, \textbf{p}^{TI}(T))],
    \end{flalign}
where $D$ denotes the pool of image-text pairs, $\textbf{y}^{IT}_V$ and $\textbf{y}^{TI}_T$ are $M$-dimensional binary vectors encoding ground-truth similarity, and $H(\cdot, \cdot)$ refers to the cross-entropy function.

\noindent\textbf{Image-Text Matching (ITM)} aims to predict whether an image-text pair is matched. 
The token embedding $x_f^{CLS}$ of the fusion encoder predicts the binary classification probability $\textbf{p}^{ITM}$. The ITM loss is defined as
    \begin{flalign}
        \loss_{ITM}=\mathbb{E}_{V,T}[H(\textbf{y}^{ITM}, \textbf{p}^{ITM}(V,T))],
    \end{flalign}
where $\textbf{y}^{ITM}$ is a binary vector indicating the matching pairs.

\noindent\textbf{Masked Language Modeling (MLM)} predicts the masked tokens in a sentence given an image and the unmasked textual tokens in the same sentence. 15\% masking probability is set. The MLM loss~\cite{devlin2018bert} is denoted as $\loss_{MLM}$. 

\noindent\textbf{Intra-Modal Contrastive (IMC)} aims to differentiate the semantics between the positive and negative pairs within the same modality \cite{yang2022vision}, i.e. image-image and text-text pairs with similarities:
    \begin{flalign}
        p_m^{II}(V)=\frac{{e^{s(V,V_m)/\tau}}}{\sum_{i=1}^M e^{s(V, V_i)/\tau}},\;
        p_m^{TT}(T)=\frac{{e^{s(T,T_m)/\tau}}}{\sum_{i=1}^M e^{s(T, T_i)/\tau}}.
    \end{flalign}
The IMC objective is defined as
    \begin{flalign}
        \loss_{IMC}=\frac{1}{2} \mathbb{E}_{V, T \sim D}[H(\textbf{y}^{II}_V, \textbf{p}^{II}(V)) + H(\textbf{y}^{TT}_T, \textbf{p}^{TT}(T))],
    \end{flalign}
where $\textbf{y}^{II}_V$ and $\textbf{y}^{TT}_T$ indicate whether the pair is matched or not. It is worth noting that $\loss_{IMC}$ encourages the model to retain the uni-modal semantics provided by the pre-trained weights of the vision and textual encoders, complementing $\loss_{ITC}$, $\loss_{ITM}$, and $\loss_{MLM}$, all of which promote multi-modal alignments.
\section{Experiments}

\subsection{Settings}

Our vision encoder is instantiated by ViT \cite{dosovitskiy2020image} and initialized with DEiT \cite{touvron2021training} or BEiT-B/16 \cite{bao2021beit} pre-trained weights. The textual encoder is initialized with BERT-Base \cite{devlin2018bert}. The proposed PiTL W-VLP model is pre-trained on three subsets IN1K, IN6K, and IN14K created from ImageNet21K. 
%Note that IN6K and IN14K are created in this work and will be further released for public access. 
Note that IN6K and IN14K are created in this work.
Specifically, IN14K contains IN6K which also contains IN1K samples. Each prompt, out of the nine shown in Table~\ref{table:prompt_example}, generates five responses for a category. Individual prompts are created for multiple synonyms, e.g. \textit{snorkeling} and \textit{snorkel\_diving} under the same category. Statistics of IN1K, IN6K, IN14K along with other datasets, e.g. CC3M, BookCorpus (BC) \cite{zhu2015aligning}, and VL-Full \cite{wang2022vlmixer} used by other VLP methods, are shown in Table~\ref{table:retrieval_results} under the Pre-training Corpus column. We assess the pre-training quality on I2T and T2I with MSCOCO-5K and Flickr30K~\cite{ordonez2011im2text}. The retrieval models are evaluated on recall at rank K (R@K).

\begin{table*}[htbp]
\caption{Results of image-text retrieval tasks on MSCOCO-5K and Flickr30K test splits. 
$^*$ denotes the results reported by~\cite{wang2023learning}.
$\dagger$ denotes BEiT~\cite{bao2021beit} self-supervised pre-trained on the images of ImageNet21K. $\ddagger$ denotes BEiT$\dagger$ intermediate fine-tuned on the images and labels of ImageNet1K. $\star$ denotes BEiT$\dagger$ intermediate fine-tuned on the images and labels of ImageNet21K.
}
\label{table:retrieval_results}

\begin{minipage}{\linewidth}

\begin{center}

\scalebox{0.7}{\setlength\tabcolsep{2.5pt}
\begin{tabular}{
p{2.5cm}|
p{3.5cm}<{\centering}|
p{2.3cm}<{\centering}|
p{2cm}<{\centering}
p{2cm}<{\centering}|
p{0.8cm}<{\centering}
p{0.8cm}<{\centering}
p{0.8cm}<{\centering}
p{0.8cm}<{\centering} 
p{0.8cm}<{\centering}
p{0.8cm}<{\centering}|
p{0.8cm}<{\centering} 
p{0.8cm}<{\centering}
p{0.8cm}<{\centering}
p{0.8cm}<{\centering}
p{0.8cm}<{\centering}
p{0.8cm}<{\centering}
}
\hline
\multirow{3}*{\textbf{VLP}} &  \multirow{3}*{Supervision}   &  \multirow{3}*{Visual Init}  &  \multicolumn{2}{c|}{Pre-training Corpus}   & \multicolumn{6}{c|}{MSCOCO-5K} &  \multicolumn{6}{c}{Flickr30K}  \\
 &  &   &   \multirow{2}*{\# of Images}  & \multirow{2}*{\# of Texts}  &  \multicolumn{3}{c}{I2T} & \multicolumn{3}{c|}{T2I} &  
 \multicolumn{3}{c}{I2T} & \multicolumn{3}{c}{T2I} \\
  &  &   &  &  &  R@1 &  R@5  &  R@10  &  R@1 &  R@5  &  R@10  &  R@1 &  R@5  &  R@10  &  R@1 &  R@5  &  R@10  \\
\hline
ViLT~\cite{kim2021vilt}   &  \multirow{3}*{\makecell[c]{image-level annotations \\ + aligned VL pairs}}   & ViT-B/32 &   4M & 4M  & 61.5  & 86.3 & 92.7 &  42.7 & 
 72.9  & 83.1 &   83.5  & 96.7  & 98.6  & 
  64.4  & 88.7 &  93.8    \\
ALBEF~\cite{li2021align}   &   &  ViT-B/16 (DeiT) & 4M & 4M  &  73.1  & 91.4 & 96.0 & 56.8 & 81.5 & 89.2  &  94.3  & 99.4  & \textbf{99.8}  & 
 82.8  & \textbf{96.7}  & 
 98.4  \\
TCL~\cite{yang2022vision}  &  & ViT-B/16 (DeiT)  & 4M & 4M  &  \textbf{75.6}  & \textbf{92.8} & \textbf{96.7}  & \textbf{59.0} & 
 \textbf{83.2} & \textbf{89.9}  &  \textbf{94.9} & 
 \textbf{99.5}  & \textbf{99.8} & 
 \textbf{84.0} & \textbf{96.7} & 
 \textbf{98.5} \\
\hline
\textbf{W-VLP }  &   &    &    &  &  &  &  \\
\hline
U-VisualBERT~\cite{li2020unsupervised}   &  \multirow{3}*{\makecell[c]{image-level + object-level \\ annotations}}  & BERT-Base  &  CC: 3M  & CC+BC: 5.5M & --  & -- & --  & --  & --  & --   &  67.8$^*$   & 90.7$^*$ &  94.9$^*$    &  55.4  &  82.9  & 89.8 \\
WFH~\cite{wang2023learning}   &    &   BERT-Base &  CC: 3M  & CC: 3M & --  & -- & --  & --  & --  & --   &  \textbf{72.0}  & \textbf{91.3} &  \textbf{95.6}  & 56.4  & 83.2 & 89.9 \\
E2E-UVLP~\cite{chen2022end}   &   &  Swin-B/32  &  CC: 3M  & CC: 3M  & --  & -- & --  & --  & --  & --   &  --  & -- &  --  & \textbf{66.4}  & \textbf{89.7} & \textbf{94.1} \\
VLMixer~\cite{wang2022vlmixer}  &  -- & BERT-Base  &  --  & --  & 57.4  &  84.0  &  91.6  & 44.0 & 74.1 & 84.1  &  -- & -- & -- & -- & -- & --  \\
VLMixer~\cite{wang2022vlmixer} &  \multirow{2}*{\makecell[c]{image-level + object-level \\ annotations}} &  BERT-Base  &  CC: 3M  & CC: 3M & 62.2 & 86.3  &  92.8  &  47.4  & 76.2  & 85.4  & --  & -- & --  & --  & --  & --  \\
VLMixer~\cite{wang2022vlmixer}  &   &  BERT-Base  & VL-Full: 5.9M   & VL-Full: 22.4M  &  \textbf{64.8}  &  \textbf{88.6}  &  \textbf{94.2}  & \textbf{50.1}  & \textbf{78.4}  & \textbf{86.9}  & --  & -- & --  & --  & --  & -- \\
\hline
PiTL & --  &  BEiT-B/16$\dagger$  & --  &  --  &  58.0  &  84.2  &  91.6  &  42.3  &  71.2  & 80.9 & 64.9 & 88.8 & 93.9 & 48.2 &  75.1  &  82.6   \\
PiTL  & \multirow{3}*{\makecell[c]{image-level \\ annotations}}  &  BEiT-B/16$\dagger$ & IN1K: 1.3M & IN1K: 45K &  63.0 &  86.9 &  93.3 & 46.9  & 75.6  & 84.8  & 78.1 & 95.9  &  98.2 &  62.8  & 87.1  & 91.7   \\
PiTL  &   &  BEiT-B/16$\dagger$ & IN6K: 4.5M & IN6K: 0.4M & 63.7 & 88.7  &  94.1  &  48.6  &  76.8 & 85.7 
 &  81.8  &  96.4  &  98.8  &  65.6  &  87.9  &  92.8  \\
PiTL  &  &  BEiT-B/16$\dagger$ & IN14K: 9M & IN14K: 1M   &  \textbf{67.4}  &  \textbf{90.2}  &  \textbf{95.2}  & \textbf{51.3}  & \textbf{78.6}  &  \textbf{86.9} & \textbf{83.9}  &  \textbf{97.5}  &  \textbf{98.6}  &  \textbf{68.9}  &  \textbf{89.8}   &  \textbf{93.8}  \\
\hline
PiTL &  -- &  ViT-B/16 (DeiT)  & --  &  --  & 58.9  & 85.0  &  91.8   
 &  44.1  &  73.1  &  81.9  & 68.6  &  89.7  & 94.6 
 &  53.7  &  79.4  &  85.4 \\
PiTL  &  \multirow{3}*{\makecell[c]{image-level \\ annotations}}  &  ViT-B/16 (DeiT) & IN1K: 1.3M  & IN1K: 45K  & 62.1  &  86.6 & 93.1  & 46.7 & 75.8 &  84.7 &   74.1 & 93.4 &	96.4 &	61.0 &	85.3 &	90.6 \\
PiTL  &   &  ViT-B/16 (DeiT) & IN6K: 4.5M & IN6K: 0.4M &  65.6 &  88.5 &  94.5  &  49.5  & 77.4  &  86.1
 &  80.1 & 95.2  &  97.8  &  66.6  &  88.7  &  92.7  \\
PiTL  &   &  ViT-B/16 (DeiT) & IN14K: 9M & IN14K: 1M  & \textbf{68.2}  & \textbf{90.4} & \textbf{95.1} & \textbf{51.9}
 & \textbf{78.7}  & \textbf{86.7}  &  \textbf{84.3} &  \textbf{97.3} & \textbf{99.0}  &  \textbf{70.3}  &  \textbf{91.0}  &  \textbf{94.3}  \\
\hline
PiTL & --  &  BEiT-B/16$\ddagger$  & --  &  --  &  62.1 &  86.7 &  93.3  &  46.2  & 75.2  & 84.3  &  75.7  &  95.2  &  97.8  &  59.8  &  84.9  &  89.9  \\
PiTL  &  \multirow{3}*{\makecell[c]{image-level \\ annotations}} &  BEiT-B/16$\ddagger$ & IN1K: 1.3M & IN1K: 45K & 63.5 & 88.3  & 93.9  & 48.3  & 76.8  & 85.4  &   79.8 & 95.7  & 98.0  & 64.7  & 88.4  &  92.5    \\
PiTL  &   &  BEiT-B/16$\ddagger$ & IN6K: 4.5M & IN6K: 0.4M &  66.0 &  88.9 &  94.6  &  49.8  & 78.2  &  86.4
 &  82.9  &  96.6  &  98.9 
 &  67.9  &  89.6  &  93.9   \\
PiTL  &   &  BEiT-B/16$\ddagger$ & IN14K: 9M & IN14K: 1M  &  \textbf{69.8}  &  \textbf{91.0}  & \textbf{95.7}   & \textbf{53.9}  & \textbf{79.8}   &  \textbf{87.8}  &  \textbf{88.0}  &  \textbf{98.7}  &  \textbf{99.5}  &  \textbf{74.6}  &  \textbf{92.4}  &  \textbf{95.6}  \\
\hline
PiTL & --  &  BEiT-B/16$\star$  &  --  & -- &  67.5  & 90.0  &  94.7 
 &  51.0  &  78.9  & 87.3 & 81.6  &  96.4  & 98.7 & 67.0  & 88.9  &  92.7 \\
PiTL  &  \multirow{3}*{\makecell[c]{image-level \\ annotations}} &  BEiT-B/16$\star$  & IN1K: 1.3M & IN1K: 45K    &  69.1  &  90.4  &  95.2  &  52.8  & 79.4  & 87.4  &  
 86.8  & 97.6  & 99.3  & 72.3  & 91.3  & 95.1 \\
 PiTL  &    &  BEiT-B/16$\star$  &  IN6K: 4.5M & IN6K: 0.4M  & 70.4 &  91.0  & 95.6    & 53.5  & 80.0  & 87.6
   & 87.2  & 98.4  & 99.3 &  73.4 & 92.4 & 95.6 \\
 PiTL  &   &  BEiT-B/16$\star$  & IN14K: 9M &  IN14K: 1M &  \textbf{71.9}   &  \textbf{92.2}   &  \textbf{96.4}  &  \textbf{55.0}  & \textbf{80.8}  & \textbf{88.2}  &  \textbf{90.7} & \textbf{98.7} & \textbf{99.5}
   & \textbf{76.2} & \textbf{93.5}  & \textbf{95.9}   \\
   
\hline
\end{tabular}}
\end{center}

\end{minipage}
\end{table*}

\subsection{Quantitative Results}

Table~\ref{table:retrieval_results} shows the main results of PiTL and the comparisons against the state-of-the-art VLP and W-VLP on the I2T and T2I tasks. 

\noindent\textbf{Effects of Initialization and Dataset Sizes}.
We initialize the image encoder with weights pre-trained with no supervision (i.e. the self-supervised BEiT$\dagger$) and with image-level supervision (i.e. ViT, BEiT$\ddagger$, and BEiT$\star$). The best performances are obtained with BEiT$\star$, whose weights contain the strongest visual semantics, on IN1K to IN14K. PiTL's results steadily improve with more images and descriptions. 
Compared to models without pre-training, BEiT$\dagger$ pre-trained on IN1K has larger improvements in R@1 than the other initializations, i.e. 8.6\% for I2T and 10.8\% for T2I R@1 on MSCOCO-5K.

\noindent\textbf{PiTL with BEiT-B/16$\dagger$ vs. W-VLP Works}.
To purely assess the generated image-text pairs, models initialized with self-supervised BEiT-B/16$\dagger$ weights are mainly benchmarked. VLMixer starts out as a better model than PiTL when both are not pre-trained on any image-text pairs. However, once pre-trained, PiTL appears to be strongly competitive. For instance, PiTL pre-trained on IN1K outperforms VLMixer pre-trained on CC3M, across all the I2T metrics, with fewer images and texts. Pre-trained on IN6K and IN14K, PiTL is strongly competitive and better than VLMixer pre-trained on VL-Full. On Flickr30K, PiTL is also comparable to E2E-UVLP with more images and fewer texts.

\noindent\textbf{PiTL Models Pre-trained on IN14K vs. W-VLP Works}.
While empowered by the pre-trained weights, i.e. ViT, BEiT-B/16$\ddagger$, and BEiT-B/16$\star$, PiTL is strongly favored over other W-VLP works across all the metrics. We stress that obtaining pre-trained weights requires far less amount of supervision, i.e. 14M labels for 14M images versus 10+ object annotations per image, compared to an effectively pre-trained OD.

\vspace{-0.3cm}
\subsection{Ablation Studies}\label{sec:ablations}
We study the retrieval performances on Flickr30K on the types of prompts (\textbf{P1}-\textbf{P9} in Table~\ref{table:prompt_example}) and the effect of the paired images and descriptions generated by PiTL.

\noindent\textbf{On Effects of Prompts P1-P9}.
We dissect the effect contributed by prompts \textbf{P1}-\textbf{9} to the retrieval performance. We pre-train models initialized with BEiT-B/16$\star$ on descriptions from each prompt. Table~\ref{table:prompt_type} shows that, with all the descriptions available, the models achieve the best R@1 scores for both retrieval tasks. On I2T, 
\textbf{P9} results in the best average recall over R@{1,5,10}. We reckon that \textbf{P9} is a more open-ended question than the others, which focus specifically on appearances, relations,  etc. The open-endedness could result in more diverse descriptions and hence better I2T result. On T2I, \textbf{P1}-\textbf{4} usually yield better T2I results than the other prompts. This seems to be expected since \textbf{P1-4} are more visually relevant.

\begin{table}[t]
\caption{Effects of each type of prompt of Table~\ref{table:prompt_example} on retrieval tasks. VA is short for visual appearances. }
\label{table:prompt_type}
%Please refer to Table~\ref{table:prompt_example} for the prompt examples.

\begin{minipage}{\linewidth}

\begin{center}

\scalebox{0.675}{
\setlength\tabcolsep{4pt}
\begin{tabular}{
p{2.6cm}<{\raggedright}|
p{0.7cm}<{\centering}
p{0.7cm}<{\centering}
p{0.7cm}<{\centering}
p{0.7cm}<{\centering}|
p{0.7cm}<{\centering} 
p{0.7cm}<{\centering} 
p{0.7cm}<{\centering}
p{0.7cm}<{\centering}| 
p{0.7cm}<{\centering} 
}
\hline
 & \multicolumn{4}{c|}{I2T} & \multicolumn{4}{c|}{T2I} & Overall \\
Types & R@1 &  R@5  &  R@10  &  AvgR  &  R@1 &  R@5  &  R@10  & AvgR  &  AvgR  \\
\hline
\textbf{P1}: colors  & 85.4 & 97.7 & 98.9   & 94.0  &  71.7  & 91.6  & 94.9 &  86.0 &  90.0   \\   
\textbf{P2}: shapes  & \underline{86.1}  & 97.7 & 98.9  &  94.2 & 70.7  & \underline{92.0}  &   95.2 & 86.0  & 90.1  \\ 
\textbf{P3}: textures  & 84.9  & 97.3  &  \underline{99.4}  &  93.9
  &  \underline{72.0}  &  \underline{92.0}  & \underline{95.4} & \textbf{86.5}  &  90.2  \\ 
\textbf{P4}: summarized VA  & \underline{86.1}  &  97.3  & 98.9 & 94.1  & 70.5  & \textbf{92.1}  & \underline{95.4}  & 86.0  & 90.0   \\ 
\textbf{P5}: scenes  & 85.7 & \underline{98.2}  &  99.3  & \underline{94.4}  & 71.7  &  91.5  &  \textbf{95.6} & \underline{86.3}  & \underline{90.3}  \\ 
\textbf{P6}: relations  & 84.8  & \textbf{98.3}  & 99.0 & 94.0  & 70.3  &  91.1  & 95.1 & 85.5  & 89.8   \\ 
\textbf{P7}: places  & 83.7  &  97.5  & 99.3 & 93.5  & 70.0 &  91.4  & 95.1 &  85.5 & 89.5  \\ 
\textbf{P8}: activities  & 85.2 &  98.1 & 99.0  & 94.1  & 71.4 & 91.4  & 95.1 &  86.0 & 90.0  \\ 
\textbf{P9}: first-person  & 85.6  &  \underline{98.2}  & \textbf{99.5} &  \underline{94.4} & 70.0  & 91.5  & 95.2 & 85.5  & 90.0   \\ 
 All & \textbf{86.8}  & 97.6  & 99.3 & \textbf{94.6}  & \textbf{72.3}  & 91.3  & 95.1 & 86.2  &  \textbf{90.4}   \\
\hline
\end{tabular}
}
\end{center}
\end{minipage}
\vspace{-0.4cm}
\end{table}

\noindent\textbf{On the Effect of Increased Uni-modal Shuffled Sources.} To study whether the improvements actually come from the proposed prompt-based weak supervision, rather than from the pure increase in the number of images and texts, we pre-train the models (initialized with BEiT-B/16$\star$) on IN1K and IN6K with shuffled images and descriptions. Indeed, on Flickr30K, the models degrade in R@1 from 60.7 to 56.1 for I2T, and from 48.2 to 42.7 for T2I, when pre-trained on shuffled IN1K and IN6K, respectively. This concludes the effectiveness of PiTL-generated pairs for VLP.

\iffalse
\begin{table}[htbp]
\begin{minipage}{\linewidth}

\begin{center}

\scalebox{0.7}{\setlength\tabcolsep{6pt}
\begin{tabular}{p{3cm}
p{0.8cm}<{\centering}
p{0.8cm}<{\centering}
p{0.8cm}<{\centering}
p{0.8cm}<{\centering}
p{0.8cm}<{\centering}
p{0.8cm}<{\centering} 
p{0.8cm}<{\centering} 
}
\toprule
 &  & \multicolumn{3}{c}{Text Retrieval} & \multicolumn{3}{c}{Image Retrieval}  \\
Method & PT  & R@1 &  R@5  &  R@10  &  R@1 &  R@5  &  R@10   \\
\midrule
  &  IN1K  &  60.7  &  87.7  & 93.3  & 48.18 &  76.8  & 85.04 \\ 
  &  IN5K  &  56.1  &  84.0  &  92.3  &  42.74 &  73.86  &  82.76 \\  
\bottomrule
\end{tabular}}
\end{center}
\caption{Results . }
\label{table:random_pair}

\end{minipage}
\end{table}
\fi

\section{Conclusion}

In this work, we proposed Prompts-in-The-Loop (PiTL), a weakly-supervised method to pre-train VL-models for cross-modal retrieval tasks. Without object-level supervision based on OD, our PiTL comes with image-level supervision from images and their categories described by large language models.
Retrieval results on the cross-modal datasets demonstrated the effectiveness of PiTL.

%%
%% The acknowledgments section is defined using the "acks" environment
%% (and NOT an unnumbered section). This ensures the proper
%% identification of the section in the article metadata, and the
%% consistent spelling of the heading.
\begin{acks}
This work is supported by the Academy of Finland in project 345791.
We acknowledge the LUMI supercomputer, owned by the EuroHPC Joint Undertaking, hosted by CSC and the LUMI consortium.
\end{acks}

%%
%% The next two lines define the bibliography style to be used, and
%% the bibliography file.
\bibliographystyle{ACM-Reference-Format}
\bibliography{sample-sigconf}

%%
%% If your work has an appendix, this is the place to put it.

\end{document}